\documentclass[prl,twocolumn,floatfix,superscriptaddress,nofootinbib,nobalancelastpage]{revtex4-1}
\usepackage{dcolumn,amsmath}
\usepackage{graphicx}
\usepackage{color}
\usepackage{bm}
\usepackage{hyperref} 
\usepackage{physics,textcomp}
\usepackage{changepage}
\usepackage{lipsum}
\usepackage{subcaption}
\usepackage{natbib}
\usepackage{array}

\newcommand{\Ti}{$\mathcal{T}$}
\newcommand{\Par}{$\mathcal{P}$}
\newcommand{\eEDM}{{\em e}EDM}

\newcommand{\ecm}{\ensuremath{e {\cdotp} {\rm cm}}}

\newcommand{\de}{d_\mathrm{e}}
\begin{document}
\title{Magnetic quadrupole moment of $^{175}$Lu and parity-violating polarization degree of levels in 
$^{175}$LuOH$^+$}

\author{Igor Kurchavov}
\email{kurchavov\_ip@pnpi.nrcki.ru}
\affiliation{Petersburg Nuclear Physics Institute named by B.P.\ Konstantinov of National Research Center ``Kurchatov Institute'' (NRC ``Kurchatov Institute'' - PNPI), 1 Orlova roscha mcr., Gatchina, 188300 Leningrad region, Russia}

\author{Daniel Maison}
\email{daniel.majson@gmail.com, maison\_de@pnpi.nrcki.ru}
\affiliation{Petersburg Nuclear Physics Institute named by B.P.\ Konstantinov of National Research Center ``Kurchatov Institute'' (NRC ``Kurchatov Institute'' - PNPI), 1 Orlova roscha mcr., Gatchina, 188300 Leningrad region, Russia}
%\affiliation{Saint Petersburg State University, 7/9 Universitetskaya nab., St. Petersburg, 199034 Russia}
\homepage{http://www.qchem.pnpi.spb.ru    }

\author{Leonid Skripnikov}
\email{skripnikov\_lv@pnpi.nrcki.ru,leonidos239@gmail.com}
\affiliation{Petersburg Nuclear Physics Institute named by B.P.\ Konstantinov of National Research Center ``Kurchatov Institute'' (NRC ``Kurchatov Institute'' - PNPI), 1 Orlova roscha mcr., Gatchina, 188300 Leningrad region, Russia}
\affiliation{Saint Petersburg State University, 7/9 Universitetskaya nab., St. Petersburg, 199034 Russia}

\author{Matt Grau}
\email{mgrau@odu.edu}
\affiliation{Department of Physics, Old Dominion University, Norfolk, VA 23529}

\author{Alexander Petrov}
\email{petrov\_an@pnpi.nrcki.ru}
\affiliation{Petersburg Nuclear Physics Institute named by B.P.\ Konstantinov of National Research Center ``Kurchatov Institute'' (NRC ``Kurchatov Institute'' - PNPI), 1 Orlova roscha mcr., Gatchina, 188300 Leningrad region, Russia}
\affiliation{Saint Petersburg State University, 7/9 Universitetskaya nab., St. Petersburg, 199034 Russia}

\date{Received: date / Revised version: date}
\begin{abstract}{
 The calculation of the parity-violating polarizations in the external electric field, which are associated with the electron electric dipole moment ($e$EDM) and magnetic quadrupole moment (MQM) of the $^{175}$Lu nucleus, as well as the determination of the rovibrational structure for the $^{175}$LuOH$^+$ cation, is performed. 
 Beyond the bending of the molecule, the slight effect of the stretching of the distance between Lu and OH is taken into account.
This study is required for the preparation of the experiment and for the extraction of the $e$EDM and MQM values of $^{175}$Lu from future measurements.
} 
\end{abstract}%end of abstract
\maketitle
\section{Introduction}
%\label{intro}
The pursuit of understanding the fundamental symmetries governing the laws of physics has been a central focus in the field of theoretical and experimental physics. Among these symmetries, the invariance with respect to charge conjugation ($\mathcal{C}$), spatial parity ($\mathcal{P}$), and time reversion ($\mathcal{T}$) has long been considered crucial. However, in the latter half of the 20th century, experimental evidence emerged, confirming the violation of both $\mathcal{P}$ and the combined $\mathcal{CP}$ symmetries in weak interactions.
Nowadays, the violation of $\mathcal{CP}$ symmetry holds immense interest in the fields of cosmology and astrophysics, as it is considered one of the three fundamental conditions for baryogenesis \cite{sakharov1967violation}. As a result, the exploration of novel manifestations of symmetry violation has emerged as a prominent research area in contemporary theoretical and experimental physics \cite{Safronova:18}.

One of the approaches used to explore these phenomena is by studying the electron electric dipole moments (\eEDM) and nuclear magnetic quadrupole moments (MQM), which serve as highly sensitive probes for testing the boundaries of the Standard Model of electroweak interactions and its extensions \cite{whitepaper, Fukuyama2012,PospelovRitz2014,YamaguchiYamanaka2020,YamaguchiYamanaka2021,flambaum2014time}.
 %Experiments aiming to measure the electron EDM ($e$EDM) and other $\mathcal{T,P}$-odd effects have provided stringent constraints on theoretical parameters \cite{ACME:18}. Triatomic molecules, such as YbOH, RaOH, and LuOH$^+$, have emerged as promising candidates for precision experiments due to their unique electronic and vibrational structures, offering enhanced sensitivity to $\mathcal{T,P}$-odd effects \cite{Kozyryev:17, Isaev_2017, maison2020study}.
The search for both \eEDM and MQM has been gaining momentum. Recently the JILA group has obtained a new constraint on the \eEDM, $|\de|<4.1\times 10^{-30}$ \ecm\ (90\% confidence) \cite{JILA23}, using the $^{180}$Hf$^{19}$F$^+$ ions trapped in a rotating electric field. It further improves the latest ACME collaboration result obtained in 2018, $|d_e| \lesssim 1.1\cdot 10^{-29}\ e\cdot\textrm{cm}$ \cite{ACME:18}, by a factor of 2.4 and the first result $|\de|\lesssim 1.3\times 10^{-28}$ on the $^{180}$Hf$^{19}$F$^+$ ions  \cite{Cornell:2017} by a factor of about 32. Planned experiments aiming to measure MQM with the use of molecules offer a pathway to investigate \Par,\Ti-odd nuclear forces, quark chromo-EDMs, and other CP-violating quark interactions \cite{PhysRevLett.113.263006}.

In this paper, we focus on the $^{175}$Lu MQM in the LuOH$^+$ molecular ion, specifically in the ground rotational level of the first excited bending vibrational mode. The choice of LuOH$^+$ is motivated by its electronic structure similarities to YbOH, while demonstrating even higher sensitivity to nuclear $\mathcal{CP}$-violation effects due to the significant electric quadrupole moment of $^{175}$Lu \cite{Maison:20a}.

In a polar molecule with a heavy element, the $\mathcal{T,P}$-violating energy shifts induced by $e$EDM and MQM are
\begin{equation}
\Delta E_{\mathcal{P},\mathcal{T}}= P_e E_{\rm eff} d_e + P_MW_MM,
\label{shift}
\end{equation}
where $d_e$ is the value of the $e$EDM, $M$ is the value of the MQM, factors $E_{\rm eff}$ and $W_M$ are determined by the electronic structure of the molecule, $P_e$ and $P_M$ are the corresponding ${\mathcal{P},\mathcal{T}}$-odd polarization coefficients.
To extract $M$ and $d_e$ from the measured energy shift $\Delta E_{\mathcal{P},\mathcal{T}} $, one needs to know $E_{\rm eff}$, $W_M$,  $P_e$ and $P_{ M}$ values. 
%LS: I've rephrase to stress that MQM and de can be separated:
%To distinguish \eEDM\ and MQM contributions at least two levels with different $P_e/P_M$ ratios have to be used. The value of $W_M$ was calculated in Ref. \cite{ Maison:20a}, $P_e$ and $E_{\rm eff}$ were calculated in Ref. \cite{Maison:22}.  
The value of $W_M$ was calculated in Ref. \cite{ Maison:20a}, $E_{\rm eff}$ was calculated in Ref. \cite{Maison:22}.
An important task is to distinguish between the two sources of symmetry violation. This can be achieved by using different molecules or different electronic states of the same molecule. However, as can be seen from Eq.~(\ref{shift}), it is also possible to utilize two or more sublevels of the same electronic state with different $P_e/P_M$ ratios. Therefore, an accurate method for estimating these ratios is necessary. In Ref. \cite{Maison:22}, we calculated the value of $P_e$ for $^{175}$LuOH$^+$. In Ref. \cite{Kurchavov2022}, a method for calculating $P_M$ was developed and applied to the $^{173}$YbOH molecule. The main objective of this study is to apply this method to calculate $P_M$ for $^{175}$LuOH$^+$. Furthermore, more precise data for $P_e$, hyperfine structure, and spectroscopic constants are obtained.

%\section{Rovibrational levels and hyperfine structure calculation details}
\section{Methods}
\begin{figure}[h]
\centering
\includegraphics[width=0.4\textwidth]{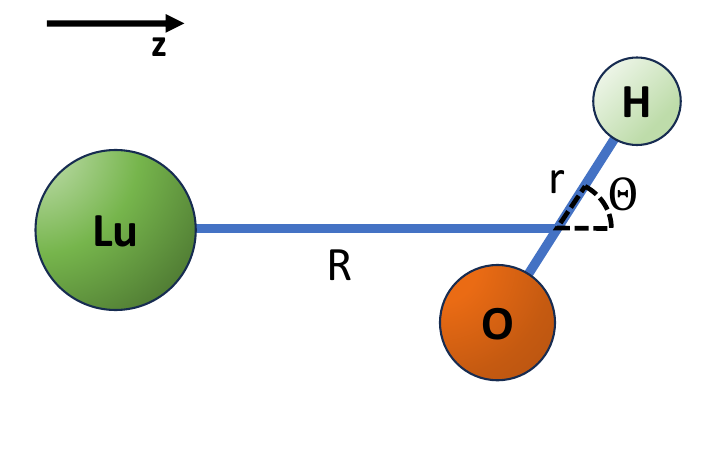}
\caption{Jacobi coordinates for the LuOH$^+$ cation.}
\label{Fig1}
\end{figure}

Following Ref. \cite{Petrov:2022}, we present the Hamiltonian in the molecular reference frame as follows:
\begin{equation}
{\rm \bf\hat{H}} = {\rm \bf\hat{H}}_{\rm mol} + {\rm \bf\hat{H}}_{\rm hfs} + {\rm \bf\hat{H}}_{\rm ext}.
\label{Hamtot}
\end{equation} 
Two approaches for the molecular Hamiltonian ${\rm \bf\hat{H}}_{\rm mol}$ are used:

\begin{equation}
\hat{\rm H}_{\rm mol}^{\rm I}=\frac{(\hat{\bf J} -\hat{\bf J}^{e-v} )^2}{2\mu R_e^2}+\frac{(\hat{\bf J}^{v})^2}{2\mu_{OH}r^2}+ V(R_e,\theta)
\label{Hmolf}
\end{equation}
and
\begin{equation}
\hat{\rm H}_{\rm mol}^{\rm II}=-\frac{\hbar^2}{2\mu}\frac{\partial^2}{\partial R^2}+\frac{(\hat{\bf J} -\hat{\bf J}^{e-v} )^2}{2\mu R^2}+\frac{(\hat{\bf J}^{v})^2}{2\mu_{OH}r^2}+ V(R,\theta),
\label{Hmolf}
\end{equation}
where
$\mu$ is the reduced mass of the Lu-OH system, $\mu_{OH}$ is the reduced mass of the OH, $\hat{\bf J}$ is the total electronic, vibrational, and rotational
angular momentum, $\hat{\bf J}^{e-v} = \hat{\bf J}^{e} + \hat{\bf J}^{v}$ is the electronic-vibrational momentum, $\hat{\bf J}^{e}$, is the electronic momentum, $\hat{\bf J}^{v}$ is the vibrational momentum,
$R$ is the distance between Lu and the center mass of OH, $R_e=1.930$ \AA~ is the corresponding equilibrium value for $R$, $r=0.954$ \AA~ is OH bond length,
 $\theta$ is the angle between OH  and the axis ($z$ axis of the molecular frame) directed from Lu to the OH center of mass and $V(R, \theta)$ is the potential energy surface obtained in the electronic structure calculations. The condition $\theta=0$ corresponds to the linear configuration where the O atom is between Lu and H ones. $R$, $r$ and $\theta$ are the so called Jacobi coordinates, see Fig.~\ref{Fig1}.

Using the Hamiltonian $\hat{\rm H}_{\rm mol}^{\rm I}$, we neglect the influence of the stretching $\nu_1$ (associated with R) and OH ligand $\nu_3$ (associated with r) vibrational modes. However, we still consider the bending modes 
$\nu_2$ (associated with $\theta$) with fixed $R$ and $r$. In this approach, the spectroscopic constants and $P_e$ coefficient in Ref. \cite{Maison:22} were calculated. Using the Hamiltonian $\hat{\rm H}_{\rm mol}^{\rm II}$, we additionally take into account the influence of the stretching mode. In this paper, we recalculated the spectroscopic constants (and obtained a new one for the stretching mode) and the $P_e$ value using $\hat{\rm H}_{\rm mol}^{\rm II}$. We also calculated the $P_M$ polarization in both approaches.

The Hamiltonian for the hyperfine interaction of electrons with Lu and H nuclei reads 
\begin{equation}
\begin{aligned}
 {\rm \bf\hat{H}}_{\rm hfs} = 
- { g}_{\rm H} {\bf \rm I^H} \cdot
 \sum_a\left(\frac{\bm{\alpha}_{2a}\times \bm{r}_{2a}}{r_{2a}^3 }\right) + \\
-{ g}_{\rm Lu}{\mu_{N}} {\bf \rm I^{Lu}} \cdot
\sum_a\left(\frac{\bm{\alpha}_a\times \bm{r}_{1a}}{{r_{1a}}^3}\right) \\
-e^2 \sum_q (-1)^q \hat{Q}^2_q({\bf \rm I^{\rm Lu}}) \sum_a \sqrt{\frac{2\pi}{5}}\frac {Y_{2q}(\theta_{1a}, \phi_{1a})}{{r_{1a}}^3}
\end{aligned}
\end{equation}
where
${ g}_{\rm Lu}$ and ${ g}_{\rm H}$ are the
 g-factors of the lutetium and hydrogen nuclei, $\bm{\alpha}_a$
 are the Dirac matrices for the $a$-th electron, $\bm{r}_{1a}$ and $\bm{r}_{2a}$ are their
 radius-vectors in the coordinate system centered on the Lu and H nuclei,
 $\hat{Q}^2_q({\bf \rm I^{\rm Lu}})$ is the quadrupole moment operator for $^{175}$Lu nucleus, $I^{\rm Lu}=7/2$, $I^{\rm H}=1/2$ are nuclear spins for $^{175}$Lu and $^{1}$H.

The Stark Hamiltonian
\begin{equation}
 {\rm \bf\hat{H}}_{\rm ext} =   -{ {\bf D}} \cdot {\bf E}
\end{equation}
describes the interaction of the molecule with the external electric field, and
{\bf D} is the dipole moment operator.

Wavefunctions, rovibrational energies, and hyperfine structure were obtained by numerically diagonalizing the Hamiltonian (\ref{Hamtot}) over the basis set of electronic-rotational-vibrational-nuclear spins wavefunctions.
\begin{equation}
 \Psi_{\Omega m\omega}P_{lm}(\theta)\Theta^{J}_{M_J,\omega}(\alpha,\beta)U^{\rm H}_{M^{\rm H}_I}U^{\rm Lu}_{M^{\rm Lu}_I}.
\label{basis0}
\end{equation}
or
\begin{equation}
 \Psi_{\Omega m\omega}\chi_{\nu_1}(R)P_{lm}(\theta)\Theta^{J}_{M_J,\omega}(\alpha,\beta)U^{\rm H}_{M^{\rm H}_I}U^{\rm Lu}_{M^{\rm Lu}_I}
\label{basis}
\end{equation}
for $\hat{\rm H}_{\rm mol}^{\rm I}$ or $\hat{\rm H}_{\rm mol}^{\rm II}$ respectively.
Here 
 $\Theta^{J}_{M_J,\omega}(\alpha,\beta)=\sqrt{(2J+1)/{4\pi}}D^{J}_{M_J,\omega}(\alpha,\beta,\gamma=0)$ is the rotational wavefunction,
 $D^{J}_{M_J,\omega}$ is Wigner function,
 $\alpha,\beta$ correspond to azimuthal and polar angles of the molecular $z$ axis (directed from Lu to the center of mass of OH group), $U^{\rm H}_{M^{\rm H}_I}$ and $U^{\rm Lu}_{M^{\rm Lu}_I}$ are the hydrogen and lutetium nuclear spin wavefunctions, $M_J$ is the projection of the molecular (electronic-rotational-vibrational) angular momentum $\hat{\bf J}$ on the 
 lab axis,  $\omega$ is the projection of the same momentum on $z$ axis of the molecular frame,
 $M^{\rm H}_I$ and $M^{\rm Lu}_I$ are the projections of the nuclear angular 
momenta of hydrogen and lutetium on the lab axis, $P_{lm}(\theta)$ is the associated Legendre polynomial, $l$ is the vibration angular momentum and $m$ is its projection on the molecular axis, $\Omega$ is the projection of the total electronic angular momentum on the molecular axis $z$ for linear configuration. $\Psi_{\Omega m\omega}$ is the electronic wavefunction (see Ref. \cite{Petrov:2022} for details).

In this  calculation functions with $\omega - m = \Omega = \pm 1/2$, $l=0-30$  and $m=0,\pm 1, \pm 2$, $J=1/2,3/2,5/2$  were included in the basis sets (\ref{basis0}) and (\ref{basis}).
The ground vibrational state $\nu_2=0$ corresponds to $m=0$,
the first excited bending mode $\nu_2=1$ to $m=\pm 1$, the second excited bending mode has states with $m=0, \pm2$ etc. A common designation $\nu_2^m$ for the bending vibrational levels will be used below. No momenta is associated with $\nu_1$ quantum number.

Provided that the {\it electronic-vibrational} matrix elements are known, the matrix elements of ${\rm \bf\hat{H}}$ between states in the basis set (\ref{basis}) can be calculated with help of the angular momentum algebra \cite{LL77, Petrov:2022} mostly in the same way as for the diatomic molecules \cite{Petrov:11}. The corresponding matrix elements are taken from Ref. \cite{Maison:22}.
To calculate the $\mathcal{T,P}$-odd shifts the average value of corresponding Hamiltonians 
%pnpi
\cite{Kozlov:87}
%end pnpi
 \begin{align}
\label{hamq}
 H_{\rm MQM}  &=
 -\frac{  M}{2I^{\rm Lu}(2I^{\rm Lu}-1)}  T_{ik}\frac{3}{2r^5}\epsilon_{jli}\alpha_jr_lr_k,
 \end{align}
where $\epsilon_{jli}$ is the unit antisymmetric tensor, $\bm{\alpha}$ is the vector of Dirac matrices,
\begin{align}\label{eqaux1}
 T_{ik}=I^{\rm Lu}_i I^{\rm Lu}_k + I^{\rm Lu}_k I^{\rm Lu}_i -\tfrac23 \delta_{i,k} I^{\rm Lu}(I^{\rm Lu}+1) ,
 \end{align}

% \eEDM\ interaction is described by the Hamiltonian

\begin{eqnarray}
  H_{\rm EDM}=2d_e
  \left(\begin{array}{cc}
  0 & 0 \\
  0 & \bm{\sigma E} \\
  \end{array}\right)\ ,
 \label{Hd}
\end{eqnarray}
$\bm{E}$ is the inner molecular electric field, and $\bm{\sigma}$ are the Pauli matrices. 
\cite{Petrov:2022, Kurchavov2022} were evaluated.

%\section{Electronic structure calculation details}

For the calculation of the potential energy surface, we employed two models to account for electron correlation effects: coupled cluster with single and double excitation amplitudes (CCSD) and coupled cluster with single, double, and perturbative triple excitation amplitudes (CCSD(T)). These calculations were performed within the Dirac-Coulomb Hamiltonian framework. For Lu, we utilized the uncontracted Dyall's AE3Z basis set~\cite{gomes:2010}, while for F, we employed the aug-cc-PVTZ-DK basis set~\cite{Dunning:89, Kendall:92}. In these calculations, the correlation treatment excluded the $1s..3d$ electrons, and a virtual orbital energy cutoff of $70\ \textrm{Hartree}$ was set.

We obtained the CCSD and CCSD(T) potential surface $V(R,\theta)$ on the grid of coordinates $(R_i,\theta_k)$,
\begin{align}
\{R_i\}=1.771,           1.877,           1.930,           1.983,           2.089 \AA \\
%\{r_j\}=1.632, 1.732,\ldots 2.032\,\mathrm{a.u.}\\
\{\theta_k\}=0^\circ, 5^\circ, 10^\circ, 15^\circ, 20^\circ, 25^\circ, 55^\circ, 90^\circ, 122^\circ, 155^\circ, 180^\circ
\end{align}
The potentials were approximated by polynomial $\sum_{n=0}^{7}\sum_{m=0}^{4}{C_{nm}\theta^n R^m}$. Since the linear configuration has minimum energy, the coefficients $C_{nm}=0$ for $n=1$.

\section{Results and discussions}
\begin{figure}[t]
\centering
  \includegraphics[width=1\linewidth]{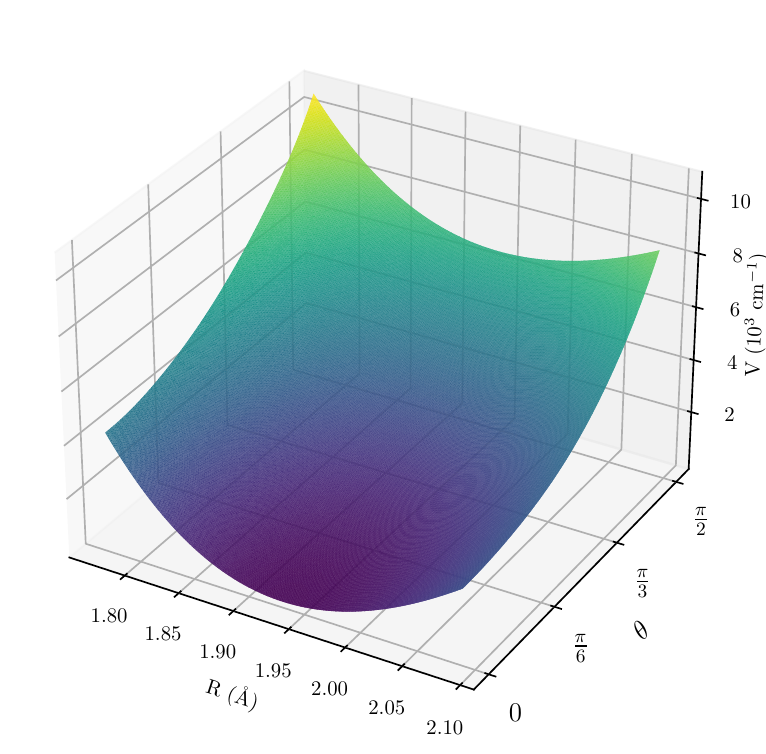}
  \caption{Potential surface $V(\theta, R)$ for CCSD(T) calculations with polynomial interpolation.}
  \label{curve}
\end{figure}

In Fig.~\ref{curve} and Table~\ref{spec}, we present the calculated potential energy surface and spectroscopic properties. It can be see that the results for the CCSD and CCSD(T) models are in close agreement with each other. Incorporating non-iterative triple cluster amplitudes leads to a decrease in vibrational energies by 4--8 cm$^{-1}$. The use of Hamiltonian $\hat{\rm H}_{\rm mol}^{\rm II}$ instead of $\hat{\rm H}_{\rm mol}^{\rm I}$ has a slightly greater impact on the vibrational energies. The final $l-$doubling value for $\nu_2=1$ of 24.5 MHz is approximately 1 MHz higher than the value obtained with a frozen $R$ variable.

\begin{figure}[t]
%\begin{center}
%\centering
\includegraphics[width=1\linewidth]{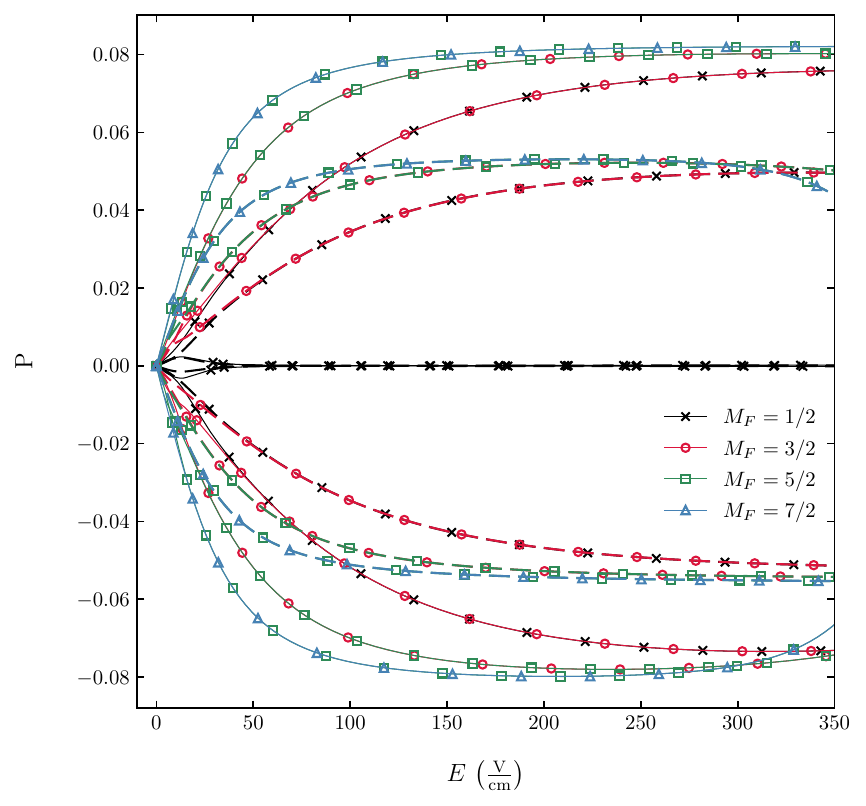}
\caption{\label{EDMMQMshift2} 
 Calculated polarizations $P_e/7$ (solid) and $P_M$ (dashed) for the selected levels (see text for details) for the different values of $M_F$ of the lowest $N=1$ rotational level of the first excited the $\nu_2=1$ bending vibrational mode of $^{175}$LuOH+ as functions of the external electric field.}
%\end{center}
\end{figure}

In Fig. \ref{EDMMQMshift2} the calculated polarizations $P_e$, $P_M$ for the selected levels of the lowest $N=1$ rotational state of the first excited $\nu_2=1$ bending vibrational mode with frozen $R$ approximation as a functions of the external electric field are given. The selected 14 levels, numbered 43–56 (see Tables \ref{spec2} and \ref{spec3}), are those which were chosen in Ref. \cite{Maison:22} for the \eEDM\ search. The corresponding energies (order of 31 850 MHz) are given in Fig. 3 panel (a) of Ref. \cite{Maison:22}. 
Using only levels numbered 43–56 is not enough, however, if nonzero MQM of $^{175}$Lu nucleus is assumed. The reason is that the ratio $P_e/P_M \sim -10.5$ is about the same for all these levels that makes it impossible to distinguish \eEDM\ and MQM contributions. 
Similarly to $P_e$ \cite{Maison:22} there are levels with close values of $P_M$. These states differ by only projection ${M^{\rm H}_I=\pm 1/2}$ which almost does not influence $P_e$ and $P_M$.

The numerical data for $P_e$, $P_M$ and hyperfine energies for all  $N=1$ levels for $E=50$ and $E=100$ V/cm are given in Tables \ref{spec2} and \ref{spec3} respectively.  To assess the influence of the stretching mode calculations with Hamiltonians $\hat{\rm H}_{\rm mol}^{\rm I}$ and $\hat{\rm H}_{\rm mol}^{\rm II}$ were performed. One can see that accounting for stretching mode leads to a decreasing of notable $P_e$, $P_M$ values up to about 5\% for $E=50$  and 4\% for $E=100$.  This is explained by increasing $l-$doubling   value describing the energy difference between levels of opposite parity at zero electric field when stretching mode is taken into account.

\begin{table}[b]
\caption{\label{spec} Calculated vibrational energy levels (${\rm cm}^{-1}$) rotational constants $B$ (${\rm cm}^{-1}$), and $l-$doubling (MHz) for the excitation modes of stretching $\nu_1=0-1$ and bending $\nu_2=0-2$ quanta of $^{175}$LuOH$^+$. Ligand mode $\nu_3$ quanta is zero in calculations.}
\begin{ruledtabular}
\begin{tabular}{cccc}
Parameter  & \parbox{2cm}{CCSD(T)\\(R-frozen)} & CCSD & CCSD(T) \\
\hline
$\nu_1=0, \nu_2=0$                   &    0     &     0      &     0        \\
$\nu_1=0, \nu_2=1$                   &   442    &    438     &    434       \\
$\nu_1=1,\nu_2=0$                    &          &    750     &    745       \\
$\nu_1=0,\nu_2=2^0$                  &   871    &    864     &    856       \\
$\nu_1=0,\nu_2=2^2$                  &   898    &     887    &    879       \\
$B(\nu_1=0,\nu_2=0)$                 &  0.2879  &   0.2874   &   0.2868     \\
$B(\nu_1=0,\nu_2=1)$                 &  0.2881  &   0.2869   &   0.2863     \\
$B(\nu_1=1,\nu_2=0)$                 &          &   0.2862   &   0.2855     \\
$B(\nu_1=0,\nu_2=2^0)$               &  0.2883  &   0.2870   &   0.2864     \\
$B(\nu_1=0,\nu_2=2^2)$               &  0.2882  &   0.2919   &   0.2912     \\
$l$-doubling $(\nu_1=0,\nu_2=1)$     &  23.5    &    24.4    &   24.5       \\
$l$-doubling $(\nu_1=0,\nu_2=2^2)$   &  0.005   &   0.012    &   0.012     \\
\end{tabular}
\end{ruledtabular}
\end{table}

Since MQM induced energy shift is proportional to $P_M$  for the MQM searches the levels with large $P_M$ values are preferred. Besides,  
 to distinguish \eEDM\ and MQM contributions, the levels with different $P_e/P_M$ ratios have to be used.
 As an example of the proposed \eEDM\ contribution exclusion scheme, let us consider the first and fifty third levels for $E=50$ V/cm.
 For the first level we have $\delta E^1 = -0.3740 E_{\rm eff} d_e + 0.0794 W_MM$; for the fifty third level $\delta E^{53} = -0.3524 E_{\rm eff} d_e + 0.0333 W_MM$.
 Then the combination $\delta E^1 - 1.0613 \delta E^{53} = 0.044 W_MM$
 is independent of \eEDM\ and can be used for MQM extraction.
 Similarly, for the electric field $E=100$ V/cm, for example, one can choose levels numbered 1 (with $\delta E^1 = -0.4715E_{\rm eff} d_e + 0.1001W_MM$ and a ratio {$P_e/P_M\approx4.71$}) and 43 (with  $\delta E^{43} = 0.5294E_{\rm eff} d_e - 0.0502 W_MM$ and the ratio equal to $10.55$). Then the obtained combination is $1.1229\delta E^1 + \delta E^{43} = 0.0622 W_MM$.  
 We note also that our choice of the levels is only an example. On the base of Tables \ref{spec2}, \ref{spec3} 
 and using formula
 \begin{equation}
 \begin{aligned}
     MW_M &= \frac{1}{\Delta}\left(P_e^x\delta E^y - P_e^y\delta E^x\right), \\
     \Delta &= P_e^xP_M^y - P_e^yP^x_M
 \end{aligned}
     % \delta E^x - \frac{P_e^x}{P_e^y}\delta E^y = \left(P_M^x - \frac{P_e^xP^y_M}{P_e^y}\right)MW_M
 \end{equation}
 one can choose alternative appropriate levels for the MQM search. Here $x$ and $y$ are numbers of chosen levels.
 Similarly the \eEDM\ contribution can be determined.

Electric field $E=100$ V/cm provides almost saturated values for $P_e$ and $P_M$.
 As it was shown in Letter \cite{PhysRevA.105.L050801} that in the $l-doubling$ structure, the $P$ value tends to reach half of the maximum value for molecules with Hund's case $b$. Calculations showed that all levels have polarizations $P_e<0.58$ and $P_M<0.12$.

% \LS{Can we comment how is it easy to access chosen levels experimentally?} 
To access these energy levels in order to perform the \eEDM\ and MQM precision measurement, trapped LuOH$^+$ ions that are initially prepared in the ground rovibrational state by either optical pumping or quantum logic spectroscopy. From this point either of the states in the pair necessary to distinguish the \eEDM\ from MQM can be populated by driving a Raman transition using a pair far-detuned infrared lasers with a difference frequency tuned to the energy of the desired state. As the pairs states with opposite projection of ${M^{\rm H}_I=\pm 1/2}$ are separated by several MHz they will be straightforward to resolve experimentally.

\section{Conclusion} 
We calculated spectroscopic constants for the lowest vibrational levels, parity-violating polarizations $P_e$ and $P_M$ associated with \eEDM\ and MQM of $^{175}$Lu energy shifts in the first excited bending mode for the $^{175}$LuOH$^{+}$ cation. The levels suitable for the MQM search are determined.

\begin{acknowledgments}
Electronic structure calculations have been carried out using computing resources of the federal collective usage center Complex for Simulation and Data Processing for Mega-science Facilities at National Research Centre ``Kurchatov Institute'', http://ckp.nrcki.ru/.

Electronic correlation calculation of the potential energy surface was supported by the Russian Science Foundation (Grant No. 19-72-10019-P (https://rscf.ru/en/project/22-72-41010/). Dirac-Hartree-Fock calculations were supported by Foundation for the Advancement of Theoretical Physics and Mathematics ``BASIS'' grant according to the Research Project No. 21-1-2-47-1.
\end{acknowledgments}
%\section{acknowledgement}
%This work is supported by the Russian Science Foundation grant No. 18-12-00227.
% BibTeX users please use
% \bibliographystyle{epj}
% \bibliography{qc}
% \onecolumn{
% bibliographystyle{apsrev}
% \newpage
% \bibliography{bibs/QCPNPI,bibs/SkripnikovLib,bibs/Titov,bibs/TitovLib,bibs/VictorFlambaum,bibs/Zakharova,bibs/Zakharova1,bibs/ACME,bibs/JournAbbr,bibs/Kaldor,bibs/Kudashov,bibs/Lomachuk,bibs/PetrovLib,bibs/qc,bibs/Maison}



\begin{table*}[p]
\caption{\label{spec2}  
The calculated energies (in MHz), polarizations for the different projections of the total angular momentum $M_F$  of the lowest $N=1$ rotational  level of the first excited $v=1$ bending vibrational mode of $^{175}$LuOH$^+$ for the value of the external electric field $E=50$ V/cm. Levels are numbered by the increasing energy. Zero energy level corresponds to the lowest energy of $N=1$ states at zero electric field. Calculation with frozen R variable are marked by (f).
} 
\begin{ruledtabular}
\begin{tabular}{rrrr rrrr| rrrr rrrr}
 \#	&   $M_F$    &     	En(f)	  &   En         &    	$P_e$(f) & 	$P_e$	  & $P_M$(f)   &     	$P_M$ &	\#  &	$M_F$&  En(f)   &	En     & $P_e$(f)& $P_e$    & $P_M$(f)  &    $P_M$      \\ \hline
 1	&   2.5   &         -5.5      &     -5.3          &   	-0.3812  &	-0.3740   &	0.0809     &   	0.0794    &	49  &	0.5  &  31851.8 & 31849.4       & 0.0014  &  0.0013  &	-0.0001  &	-0.0001    \\
 2	&   1.5   &         -5.3      &     -5.1          &   	-0.3811  &	-0.3738   &	0.0809     &   	0.0794    &	50  &	0.5  &  31873.7 & 31872.1       & -0.0013 &	-0.0012  &	0.0001   &	0.0001    \\
 3	&   1.5   &         -1.7      &     -1.7          &   	-0.2450  &	-0.2374   &	0.0520     &   	0.0504    &	51  &	0.5  &  31874.6 & 31872.9       & -0.2151 &	-0.2070  &	0.0204   &	0.0195    \\
 4	&   0.5   &         -1.6      &     -1.6          &   	-0.2436  &	-0.2359   &	0.0517     &   	0.0501    &	52  &	1.5  &  31874.6 & 31872.9       & -0.2164 &	-0.2082  &	0.0206   &	0.0197    \\
 5	&   0.5   &         -0.2      &     -0.2          &   	-0.0013  &	-0.0013   &	0.0003     &   	0.0003    &	53  &	1.5  &  31876.9 & 31875.2       & -0.3631 &	-0.3524  &	0.0345   &	0.0333    \\
 6	&   0.5   &         23.3      &     24.3          &   	0.0011   &	0.0011    &	-0.0002    &   	-0.0002   &	54  &	2.5  &  31877.0 & 31875.2       & -0.3631 &	-0.3524  &	0.0345   &	0.0333    \\
 7	&   1.5   &         24.9      &     25.8          &   	0.2448   &	0.2372    &	-0.0520    &   	-0.0503   &	55  &	2.5  &  31880.1 & 31878.3       & -0.4457 &	-0.4355  &	0.0423   &	0.0411    \\
 8	&   0.5   &         25.0      &     25.9          &   	0.2436   &	0.2360    &	-0.0517    &   	-0.0501   &	56  &	3.5  &  31880.3 & 31878.4       & -0.4457 &	-0.4355  &	0.0423   &	0.0411    \\
 9	&   2.5   &         28.7      &     29.5          &   	0.3808   &	0.3736    &	-0.0808    &   	-0.0793   &	57  &	4.5  &  32043.2 & 32042.7       & 0.3530  &  0.3435  &	-0.0477  &	-0.0475    \\
 10  &  	1.5   &         28.9      &     29.7          &   	0.3807   &	0.3734    &	-0.0808    &   	-0.0793   &	58  &	5.5  &  32043.3 & 32042.9       & 0.3528  &  0.3433  &	-0.0477  &	-0.0475    \\
 11  &  	4.5   &         118.7     &     119.3         &   	-0.3374  &	-0.3333   &	0.0703     &   	0.0694    &	59  &	3.5  &  32045.9 & 32045.4       & 0.3266  &  0.3165  &	-0.0442  &	-0.0437    \\
 12  &  	3.5   &         118.9     &     119.5         &   	-0.3374  &	-0.3333   &	0.0703     &   	0.0694    &	60  &	4.5  &  32046.1 & 32045.5       & 0.3265  &  0.3164  &	-0.0442  &	-0.0437    \\
 13  &  	3.5   &         122.2     &     122.8         &   	-0.3040  &	-0.2987   &	0.0633     &   	0.0622    &	61  &	2.5  &  32048.4 & 32047.8       & 0.2853  &  0.2750  &	-0.0386  &	-0.0380    \\
 14  &  	2.5   &         122.4     &     122.9         &   	-0.3039  &	-0.2986   &	0.0633     &   	0.0622    &	62  &	3.5  &  32048.5 & 32048.0       & 0.2853  &  0.2750  &	-0.0386  &	-0.0380    \\
 15  &  	2.5   &         125.3     &     125.8         &   	-0.2452  &	-0.2391   &	0.0511     &   	0.0498    &	63  &	1.5  &  32050.5 & 32049.9       & 0.2203  &  0.2108  &	-0.0298  &	-0.0291    \\
 16  &  	1.5   &         125.4     &     125.9         &   	-0.2449  &	-0.2387   &	0.0510     &   	0.0497    &	64  &	2.5  &  32050.6 & 32050.0       & 0.2205  &  0.2111  &	-0.0298  &	-0.0292    \\
 17  &  	1.5   &         127.5     &     127.9         &   	-0.1446  &	-0.1396   &	0.0301     &   	0.0291    &	65  &	0.5  &  32052.0 & 32051.3       & 0.1202  &  0.1141  &	-0.0162  &	-0.0158    \\
 18  &  	0.5   &         127.5     &     127.9         &   	-0.1419  &	-0.1368   &	0.0296     &   	0.0285    &	66  &	1.5  &  32052.0 & 32051.4       & 0.1241  &  0.1179  &	-0.0168  &	-0.0163    \\
 19  &  	0.5   &         128.4     &     128.7         &   	-0.0024  &	-0.0025   &	0.0005     &   	0.0005    &	67  &	0.5  &  32052.5 & 32051.8       & 0.0036  &  0.0036  &	-0.0005  &	-0.0005    \\
 20  &  	0.5   &         151.7     &     153.1         &   	0.0020   &	0.0021    &	-0.0004    &   	-0.0004   &	68  &	0.5  &  32075.7 & 32075.9       & -0.0035 &	-0.0035  &	0.0005   &	0.0005    \\
 21  &  	1.5   &         152.6     &     153.9         &   	0.1444   &	0.1394    &	-0.0301    &   	-0.0290   &	69  &	0.5  &  32076.3 & 32076.5       & -0.1200 &	-0.1139  &	0.0162   &	0.0158    \\
 22  &  	0.5   &         152.6     &     154.0         &   	0.1422   &	0.1371    &	-0.0296    &   	-0.0286   &	70  &	1.5  &  32076.3 & 32076.5       & -0.1238 &	-0.1177  &	0.0168   &	0.0163    \\
 23  &  	2.5   &         154.8     &     156.1         &   	0.2450   &	0.2388    &	-0.0510    &   	-0.0497   &	71  &	1.5  &  32077.8 & 32077.9       & -0.2198 &	-0.2103  &	0.0297   &	0.0291    \\
 24  &  	1.5   &         154.9     &     156.2         &   	0.2447   &	0.2385    &	-0.0510    &   	-0.0497   &	72  &	2.5  &  32077.8 & 32078.0       & -0.2201 &	-0.2106  &	0.0298   &	0.0291    \\
 25  &  	3.5   &         158.0     &     159.2         &   	0.3036   &	0.2984    &	-0.0632    &   	-0.0621   &	73  &	2.5  &  32080.0 & 32080.1       & -0.2847 &	-0.2743  &	0.0385   &	0.0379    \\
 26  &  	2.5   &         158.1     &     159.3         &   	0.3036   &	0.2983    &	-0.0632    &   	-0.0621   &	74  &	3.5  &  32080.1 & 32080.2       & -0.2847 &	-0.2743  &	0.0385   &	0.0379    \\
 27  &  	4.5   &         161.7     &     162.8         &   	0.3369   &	0.3329    &	-0.0702    &   	-0.0693   &	75  &	3.5  &  32082.8 & 32082.8       & -0.3257 &	-0.3156  &	0.0441   &	0.0437    \\
 28  &  	3.5   &         161.9     &     163.0         &   	0.3369   &	0.3328    &	-0.0702    &   	-0.0693   &	76  &	4.5  &  32082.9 & 32083.0       & -0.3256 &	-0.3155  &	0.0441   &	0.0437    \\
 29  &  	2.5   &         688.8     &     689.0         &   	0.0456   &	0.0440    &	-0.0091    &   	-0.0088   &	77  &	4.5  &  32085.9 & 32085.9       & -0.3520 &	-0.3424  &	0.0477   &	0.0474    \\
 30  &  	3.5   &         688.8     &     689.1         &   	0.0457   &	0.0440    &	-0.0091    &   	-0.0088   &	78  &	5.5  &  32086.0 & 32086.1       & -0.3518 &	-0.3422  &	0.0477   &	0.0474    \\
 31  &  	1.5   &         689.4     &     689.6         &   	0.0312   &	0.0300    &	-0.0062    &   	-0.0060   &	79  &	4.5  &  32319.7 & 32315.5       & 0.0083  &  0.0120  &	0.0108   &	0.0112    \\
 32  &  	2.5   &         689.4     &     689.7         &   	0.0314   &	0.0302    &	-0.0063    &   	-0.0061   &	80  &	3.5  &  32319.9 & 32315.6       & 0.0082  &  0.0119  &	0.0107   &	0.0111    \\
 33  &  	0.5   &         689.7     &     690.0         &   	0.0145   &	0.0138    &	-0.0029    &   	-0.0028   &	81  &	3.5  &  32320.3 & 32316.1       & 0.0063  &  0.0092  &	0.0083   &	0.0086    \\
 34  &  	1.5   &         689.8     &     690.0         &   	0.0162   &	0.0156    &	-0.0032    &   	-0.0031   &	82  &	2.5  &  32320.4 & 32316.2       & 0.0062  &  0.0089  &	0.0081   &	0.0084    \\
 35  &  	0.5   &         689.9     &     690.1         &   	0.0016   &	0.0015    &	-0.0003    &   	-0.0003   &	83  &	2.5  &  32320.8 & 32316.5       & 0.0044  &  0.0064  &	0.0058   &	0.0060    \\
 36  &  	0.5   &         713.4     &     714.6         &   	-0.0063  &	-0.0063   &	0.0013     &   	0.0013    &	84  &	1.5  &  32320.8 & 32316.5       & 0.0039  &  0.0057  &	0.0052   &	0.0054    \\
 37  &  	1.5   &         713.5     &     714.7         &   	-0.0195  &	-0.0192   &	0.0039     &   	0.0038    &	85  &	0.5  &  32321.0 & 32316.7       & 0.0014  &  0.0020  &	0.0018   &	0.0019    \\
 38  &  	0.5   &         713.5     &     714.7         &   	-0.0099  &	-0.0093   &	0.0020     &   	0.0019    &	86  &	1.5  &  32321.0 & 32316.8       & 0.0025  &  0.0036  &	0.0032   &	0.0034    \\
 39  &  	1.5   &         713.6     &     714.8         &   	-0.0285  &	-0.0270   &	0.0057     &   	0.0054    &	87  &	0.5  &  32321.2 & 32316.9       & 0.0007  &  0.0011  &	0.0010   &	0.0010    \\
 40  &  	2.5   &         713.6     &     714.8         &   	-0.0331  &	-0.0322   &	0.0066     &   	0.0064    &	88  &	4.5  &  32342.1 & 32338.3       & -0.0069 &	-0.0105  &	-0.0107  &	-0.0111    \\
 41  &  	2.5   &         713.7     &     714.9         &   	-0.0450  &	-0.0431   &	0.0090     &   	0.0086    &	89  &	3.5  &  32342.1 & 32338.3       & -0.0056 &	-0.0085  &	-0.0086  &	-0.0089    \\
 42  &  	3.5   &         713.8     &     715.0         &   	-0.0463  &	-0.0446   &	0.0092     &   	0.0089    &	90  &	2.5  &  32342.1 & 32338.3       & -0.0041 &	-0.0062  &	-0.0062  &	-0.0065    \\
 43  &  	2.5   &         31845.7   &     31843.5       &   	0.4474   &	0.4374    &	-0.0426    &   	-0.0414   &	91  &	1.5  &  32342.1 & 32338.3       & -0.0025 &	-0.0038  &	-0.0038  &	-0.0039    \\
 44  &  	3.5   &         31845.8   &     31843.6       &   	0.4473   &	0.4373    &	-0.0426    &   	-0.0414   &	92  &	0.5  &  32342.1 & 32338.3       & -0.0008 &	-0.0013  &	-0.0013  &	-0.0013    \\
 45  &  	1.5   &         31848.7   &     31846.4       &   	0.3642   &	0.3536    &	-0.0347    &   	-0.0335   &	93  &	3.5  &  32342.3 & 32338.5       & -0.0066 &	-0.0101  &	-0.0103  &	-0.0107    \\
 46  &  	2.5   &         31848.8   &     31846.5       &   	0.3642   &	0.3536    &	-0.0347    &   	-0.0335   &	94  &	2.5  &  32342.3 & 32338.5       & -0.0049 &	-0.0074  &	-0.0075  &	-0.0078    \\
 47  &  	0.5   &         31850.9   &     31848.6       &   	0.2156   &	0.2075    &	-0.0205    &   	-0.0196   &	95  &	1.5  &  32342.3 & 32338.5       & -0.0030 &	-0.0045  &	-0.0046  &	-0.0047    \\
 48  &  	1.5   &         31851.0   &     31848.7       &   	0.2170   &	0.2089    &	-0.0206    &   	-0.0198   &	96  &	0.5  &  32342.3 & 32338.5       & -0.0010 &	-0.0015  &	-0.0015  &	-0.0016    \\
\end{tabular}
\end{ruledtabular}
\end{table*}

\begin{table*}[p]
\caption{\label{spec3}  
The calculated energies (in MHz), polarizations for the different projections of the total angular momentum $M_F$  of the lowest $N=1$ rotational  level of the first excited $v=1$ bending vibrational mode of $^{175}$LuOH$^+$ for the value of the external electric field $E=100$ V/cm. Levels are numbered by the increasing energy. Zero energy level corresponds to the lowest energy of $N=1$ states at zero electric field. Calculation with frozen R variable are marked by (f).
} 
\begin{ruledtabular}
\begin{tabular}{rrrr rrrr| rrrr rrrr}
 \#	&   $M_F$    &     	En(f)	  &        En         &   $P_e$(f)   & 	$P_e$	     & $P_M$(f)   &   $P_M$    &	\#  &	$M_F$ &  En(f)   &	En     & $P_e$(f)& $P_e$    & $P_M$(f)  &    $P_M$      \\ \hline
 1	&  2.5       &       -16.2    &       -15.9       &  -0.4750     &   -0.4715     &   0.1009   &  0.1001    &	49  &    0.5  &  31850.8 & 31848.4 & -0.0001 & -0.0001  &  0.0000   &    0.0000    \\
 2	&  1.5       &       -16.0    &       -15.8       &  -0.4750     &   -0.4714     &   0.1009   &  0.1001    &	50  &    0.5  &  31872.6 & 31870.9 & 0.0002  & 0.0002   &  0.0000   &    0.0000   \\
 3	&  1.5       &       -6.2     &       -6.0        &  -0.3811     &   -0.3739     &   0.0809   &  0.0794    &	51  &    0.5  &  31875.8 & 31874.0 & -0.3626 & -0.3519  &  0.0345   &    0.0333   \\
 4	&  0.5       &       -6.1     &       -5.9        &  -0.3809     &   -0.3736     &   0.0809   &  0.0793    &	52  &    1.5  &  31875.8 & 31874.1 & -0.3625 & -0.3517  &  0.0345   &    0.0332   \\
 5	&  0.5       &       -0.9     &       -0.9        &  -0.0004     &   -0.0004     &   0.0001   &  0.0001    &	53  &    1.5  &  31882.8 & 31880.9 & -0.4904 & -0.4813  &  0.0466   &    0.0454   \\
 6	&  0.5       &       22.5     &       23.5        &  0.0004      &   0.0004      &   -0.0001  &  -0.0001   &	54  &    2.5  &  31882.8 & 31881.0 & -0.4903 & -0.4812  &  0.0466   &    0.0454   \\
 7	&  1.5       &       27.9     &       28.8        &  0.3807      &   0.3734      &   -0.0808  &  -0.0792   &	55  &    2.5  &  31891.1 & 31889.2 & -0.5331 & -0.5256  &  0.0506   &    0.0496   \\
 8	&  0.5       &       28.0     &       28.8        &  0.3804      &   0.3731      &   -0.0807  &  -0.0792   &	56  &    3.5  &  31891.2 & 31889.3 & -0.5330 & -0.5255  &  0.0506   &    0.0495   \\
 9	&  2.5       &       38.6     &       39.3        &  0.4741      &   0.4706      &   -0.1006  &  -0.0999   &	57  &    4.5  &  32026.8 & 32026.5 & 0.4008  & 0.3931   &  -0.0542  &    -0.0543    \\
 10  &  1.5       &       38.7     &       39.4        &  0.4741      &   0.4706      &   -0.1006  &  -0.0998   &	58  &    5.5  &  32027.0 & 32026.6 & 0.4006  & 0.3929   &  -0.0542  &    -0.0543    \\
 11  &  4.5       &       102.3    &       103.0       &  -0.3830     &   -0.3816     &   0.0798   &  0.0795    &	59  &    3.5  &  32033.0 & 32032.5 & 0.3896  & 0.3814   &  -0.0526  &    -0.0526    \\
 12  &  3.5       &       102.5    &       103.2       &  -0.3830     &   -0.3815     &   0.0798   &  0.0795    &	60  &    4.5  &  32033.1 & 32032.7 & 0.3894  & 0.3812   &  -0.0526  &    -0.0526    \\
 13  &  3.5       &       110.4    &       111.0       &  -0.3691     &   -0.3668     &   0.0769   &  0.0764    &	61  &    2.5  &  32039.0 & 32038.5 & 0.3692  & 0.3603   &  -0.0498  &    -0.0496    \\
 14  &  2.5       &       110.5    &       111.2       &  -0.3691     &   -0.3667     &   0.0769   &  0.0764    &	62  &    3.5  &  32039.1 & 32038.6 & 0.3691  & 0.3601   &  -0.0498  &    -0.0496    \\
 15  &  2.5       &       118.1    &       118.7       &  -0.3370     &   -0.3329     &   0.0702   &  0.0693    &	63  &    1.5  &  32044.6 & 32044.1 & 0.3258  & 0.3159   &  -0.0439  &    -0.0435    \\
 16  &  1.5       &       118.2    &       118.8       &  -0.3369     &   -0.3329     &   0.0702   &  0.0693    &	64  &    2.5  &  32044.7 & 32044.2 & 0.3257  & 0.3157   &  -0.0439  &    -0.0435    \\
 17  &  1.5       &       124.6    &       125.1       &  -0.2450     &   -0.2388     &   0.0510   &  0.0497    &	65  &    0.5  &  32049.2 & 32048.6 & 0.2201  & 0.2107   &  -0.0297  &    -0.0290    \\
 18  &  0.5       &       124.7    &       125.2       &  -0.2446     &   -0.2384     &   0.0509   &  0.0497    &	66  &    1.5  &  32049.2 & 32048.6 & 0.2203  & 0.2109   &  -0.0297  &    -0.0290    \\
 19  &  0.5       &       127.7    &       128.1       &  -0.0004     &   -0.0004     &   0.0001   &  0.0001    &	67  &    0.5  &  32051.2 & 32050.5 & 0.0002  & 0.0002   &  0.0000   &    0.0000    \\
 20  &  0.5       &       151.0    &       152.4       &  0.0003      &   0.0004      &   -0.0001  &  -0.0001   &	68  &    0.5  &  32074.1 & 32074.3 & -0.0002 & -0.0002  &  0.0000   &    0.0000   \\
 21  &  1.5       &       154.1    &       155.4       &  0.2448      &   0.2386      &   -0.0510  &  -0.0497   &	69  &    0.5  &  32076.2 & 32076.4 & -0.2195 & -0.2101  &  0.0296   &    0.0289   \\
 22  &  0.5       &       154.2    &       155.4       &  0.2444      &   0.2382      &   -0.0509  &  -0.0496   &	70  &    1.5  &  32076.3 & 32076.4 & -0.2197 & -0.2102  &  0.0296   &    0.0290   \\
 23  &  2.5       &       161.0    &       162.1       &  0.3365      &   0.3324      &   -0.0701  &  -0.0692   &	71  &    1.5  &  32081.2 & 32081.3 & -0.3246 & -0.3146  &  0.0438   &    0.0434   \\
 24  &  1.5       &       161.1    &       162.2       &  0.3364      &   0.3324      &   -0.0701  &  -0.0692   &	72  &    2.5  &  32081.3 & 32081.3 & -0.3245 & -0.3145  &  0.0438   &    0.0433   \\
 25  &  3.5       &       169.2    &       170.3       &  0.3684      &   0.3660      &   -0.0767  &  -0.0762   &	73  &    2.5  &  32087.6 & 32087.6 & -0.3676 & -0.3585  &  0.0496   &    0.0495   \\
 26  &  2.5       &       169.4    &       170.4       &  0.3683      &   0.3660      &   -0.0767  &  -0.0762   &	74  &    3.5  &  32087.7 & 32087.7 & -0.3674 & -0.3584  &  0.0496   &    0.0495   \\
 27  &  4.5       &       178.0    &       179.1       &  0.3821      &   0.3806      &   -0.0796  &  -0.0793   &	75  &    3.5  &  32094.7 & 32094.6 & -0.3876 & -0.3793  &  0.0524   &    0.0524   \\
 28  &  3.5       &       178.2    &       179.2       &  0.3821      &   0.3806      &   -0.0796  &  -0.0793   &	76  &    4.5  &  32094.8 & 32094.7 & -0.3874 & -0.3792  &  0.0524   &    0.0524   \\
 29  &  2.5       &       687.2    &       687.5       &  0.0794      &   0.0772      &   -0.0159  &  -0.0155   &	77  &    4.5  &  32102.1 & 32102.1 & -0.3988 & -0.3909  &  0.0540   &    0.0542   \\
 30  &  3.5       &       687.3    &       687.6       &  0.0794      &   0.0772      &   -0.0159  &  -0.0155   &	78  &    5.5  &  32102.3 & 32102.2 & -0.3985 & -0.3907  &  0.0541   &    0.0542   \\
 31  &  1.5       &       689.4    &       689.7       &  0.0585      &   0.0565      &   -0.0117  &  -0.0113   &	79  &    4.5  &  32318.7 & 32314.5 & 0.0150  & 0.0217   &  0.0190   &    0.0198   \\
 32  &  2.5       &       689.5    &       689.8       &  0.0585      &   0.0566      &   -0.0117  &  -0.0113   &	80  &    3.5  &  32318.9 & 32314.6 & 0.0150  & 0.0216   &  0.0190   &    0.0198   \\
 33  &  0.5       &       690.9    &       691.1       &  0.0311      &   0.0299      &   -0.0062  &  -0.0060   &	81  &    3.5  &  32320.9 & 32316.6 & 0.0119  & 0.0172   &  0.0154   &    0.0160   \\
 34  &  1.5       &       690.9    &       691.1       &  0.0315      &   0.0303      &   -0.0063  &  -0.0061   &	82  &    2.5  &  32321.0 & 32316.7 & 0.0118  & 0.0171   &  0.0153   &    0.0159   \\
 35  &  0.5       &       691.4    &       691.6       &  0.0003      &   0.0003      &   -0.0001  &  -0.0001   &	83  &    2.5  &  32322.4 & 32318.2 & 0.0083  & 0.0121   &  0.0109   &    0.0113   \\
 36  &  0.5       &       714.8    &       716.0       &  -0.0032     &   -0.0035     &   0.0006   &  0.0007    &	84  &    1.5  &  32322.5 & 32318.3 & 0.0082  & 0.0120   &  0.0108   &    0.0112   \\
 37  &  0.5       &       715.0    &       716.2       &  -0.0285     &   -0.0270     &   0.0057   &  0.0054    &	85  &    1.5  &  32323.4 & 32319.2 & 0.0043  & 0.0063   &  0.0057   &    0.0059   \\
 38  &  1.5       &       715.0    &       716.2       &  -0.0322     &   -0.0311     &   0.0064   &  0.0062    &	86  &    0.5  &  32323.4 & 32319.2 & 0.0040  & 0.0057   &  0.0052   &    0.0054   \\
 39  &  1.5       &       715.3    &       716.5       &  -0.0588     &   -0.0568     &   0.0117   &  0.0113    &	87  &    0.5  &  32323.8 & 32319.5 & 0.0003  & 0.0005   &  0.0004   &    0.0005   \\
 40  &  2.5       &       715.4    &       716.5       &  -0.0593     &   -0.0574     &   0.0118   &  0.0114    &	88  &    4.5  &  32344.1 & 32340.3 & -0.0126 & -0.0190  &  -0.0187  &    -0.0195    \\
 41  &  2.5       &       715.9    &       717.0       &  -0.0804     &   -0.0783     &   0.0160   &  0.0156    &	89  &    3.5  &  32344.3 & 32340.4 & -0.0120 & -0.0180  &  -0.0177  &    -0.0184    \\
 42  &  3.5       &       715.9    &       717.1       &  -0.0805     &   -0.0784     &   0.0161   &  0.0156    &	90  &    3.5  &  32344.4 & 32340.6 & -0.0110 & -0.0166  &  -0.0161  &    -0.0168    \\
 43  &  2.5       &       31833.7  &       31831.6     &  0.5366      &   0.5294      &   -0.0512  &  -0.0502   &	91  &    2.5  &  32344.4 & 32340.6 & -0.0089 & -0.0133  &  -0.0127  &    -0.0132    \\
 44  &  3.5       &       31833.8  &       31831.8     &  0.5365      &   0.5293      &   -0.0511  &  -0.0502   &	92  &    1.5  &  32344.4 & 32340.6 & -0.0054 & -0.0081  &  -0.0077  &    -0.0080    \\
 45  &  1.5       &       31841.2  &       31839.0     &  0.4928      &   0.4839      &   -0.0470  &  -0.0459   &	93  &    0.5  &  32344.5 & 32340.6 & -0.0018 & -0.0027  &  -0.0026  &    -0.0027    \\
 46  &  2.5       &       31841.3  &       31839.1     &  0.4927      &   0.4838      &   -0.0470  &  -0.0459   &	94  &    2.5  &  32344.6 & 32340.7 & -0.0091 & -0.0135  &  -0.0131  &    -0.0135    \\
 47  &  0.5       &       31847.7  &       31845.4     &  0.3638      &   0.3532      &   -0.0347  &  -0.0335   &	95  &    1.5  &  32344.6 & 32340.8 & -0.0061 & -0.0089  &  -0.0086  &    -0.0089    \\
 48  &  1.5       &       31847.7  &       31845.5     &  0.3637      &   0.3531      &   -0.0347  &  -0.0335   &	96  &    0.5  &  32344.7 & 32340.8 & -0.0021 & -0.0031  &  -0.0030  &    -0.0031    \\
\end{tabular}
\end{ruledtabular}
\end{table*}

\end{document}